\documentclass{article}
\usepackage[cp1251]{inputenc}
\usepackage[english]{babel}

\usepackage{amsmath}
\usepackage{amssymb}

\usepackage[dvips]{graphicx}

\newcommand{\N}{{\mathbb N}}
\newcommand{\vphi}{\varphi}

\newtheorem{proposition}{\textsc{Proposition}}
\newtheorem{remark}{\textsc{Remark}}
\newtheorem{theorem}{\textsc{Theorem}}

\title{Application of the Abel Equation of the 1st kind to an inflation analysis of non-exactly solvable cosmological models}
\author{Artyom V. Yurov, Anna V. Yaparova, Valerian A. Yurov}
\date{}

\begin{document}

\maketitle

\begin{abstract}
In this paper we revisit the relationship between the Einstein--Friedman and the Abel equations to demonstrate how it might be applied to the inflationary analysis in a flat Friedman universe filled with a real-valued scalar field. The analysis is performed for three distinct cases of polynomial potentials. As a result of a numeric integration of Abel equation, the necessary and sufficient conditions for both slow-rolling and inflation proper are estimated with respect to the initial value of the field. In addition, the relationship between the slow-rolling condition and the inflation is ascertained.
\\
\\
\textbf{Keywords:} cosmology, scalar field, inflation, Friedman equations, Abel equation.\\
\end{abstract}
\section{Introduction}
In early 1980s A.~Linde proposed the chaotic inflation scenario \cite{Linde_1983, Linde_86}. It was a development of a new inflationary universe scenario \cite{Linde_82,Albrecht_Steinhardt}, whose predecessors in particular have been the model, introduced by A.\,A.~Starobinsky \cite{Starobinsky_1979,Starobinsky_1980,Barrow_Ottewill}  and the inflationary scenario of A.~Guth \cite{Guth_81, Hawking_Moss_Stewart, Guth_Weinberg}. The monograph \cite{Linde} studied the inflation for the simplest model of a homogeneous and isotropic universe filled with the real-valued scalar field $\vphi$ with the Lagrangian
\begin{equation}
L=\frac{1}{2}\dot{\vphi}^2-V(\vphi).
\label{lagrangian}
\end{equation}
For this model the Einstein-Friedman equations take form
\begin{equation}
H^2+\frac{k}{a^2}=\frac{8\pi}{3\text{M}_{\text{P}}^2}\left(\frac{1}{2}\dot{\vphi}^2+V(\vphi)\right),
\label{Friedman_H2}
\end{equation}
\begin{equation}
\ddot{\vphi}+3H\dot{\vphi}+d V(\vphi)/ d \vphi=0,
\label{Friedman_ddot_phi}
\end{equation}
where $k=0,\;+1\;,-1$ correspond to either flat, closed or open Friedman universe, $a=a(t)$ is a scale factor, and $H$ is the Hubble parameter; the equations here and throughout the rest of this paper make use of a ``natural'' unit system in which $\hbar=c=1$ and $G=\text{M}_{\text{P}}^{-2}$.\par
In order to get a good approximation of the inflationary solution, it is quite customary to resort to the so-called slow-rolling approximation. It imposes the condition that the changes in a scalar field are sufficiently slow for the following inequality to hold:  $\dfrac{1}{2}\dot{\vphi}\ll|V(\vphi)|$. If we also assume that the scale factor $a(t)$ changes fast enough for $H \gg \dfrac{k}{a^2}$ while the change of $\dot{\vphi}$ is further restricted by $\ddot{\vphi} \ll V'(\vphi)$, we'll come to the conclusion that $H^2=V(\vphi)$, i.e. that the Hubble parameter has a weak time dependence, and that the scale factor increases by the law
$$
a(t)\sim e^{Ht}.
$$
Thus, the universe expands exponentially, and inflation takes place.\par
According to the assumption \cite{Linde}, an exit from inflation occurs when the slow-rolling approximation becomes invalid, i.e. when the kinetic part of the scalar field energy density $\frac{1}{2}\dot{\vphi}^2$ becomes comparable to $|V(\vphi)|$.\par
The slow-rolling condition serves to explain the occurrence of inflation. Unfortunately, in many models it appears to offer no easy natural exit from an inflationary regime. Many papers, dedicated to the problem of existence of inflation's end for the exactly solvable cosmological models employ the so-called ``fine tuning'', i.e. a model-specific accurate adjustment of the model parameters \cite{Barrow_1994,Maartens_Taylor_Roussos,Yurov_2001,Yurov_Vereshchagin}.\par
Some examples of models that have a natural exit from inflation and do not require a fickle aid of a ``fine tuning'' have been provided in \cite{Yurov_Vereshchagin}, although the potentials used therein do not necessarily correspond to any (currently known) particle theory. Besides, the definition of a universe evolution given there might be incomplete.\par
In this article we will expand the results of paper \cite{Yurov_Yurov}, where we suggested a new way to solve Friedman equations~(\ref{Friedman_H2}),~(\ref{Friedman_ddot_phi}) by means of reduction to Abel equation of the 1st kind (the process of reduction is considered in detail in section~\ref{Connection}). We will show how to use this reduction for the task of an inflation's analysis, when applied for the model of a flat Friedman universe filled with a real-valued scalar field with a potential
\begin{equation}
V(\vphi)=\frac{m^2\vphi^2}{2}+\frac{\lambda\vphi^4}{4},
\label{V}
\end{equation}
where $\lambda=10^{-14}$, $m^2=\lambda\text{M}_{\text{P}}^2$ according to \cite{Linde}. The particular goals of the paper can be summed up as follows:
\begin{itemize}
\item to ascertain the possible necessity and sufficiency properties of the slow-rolling conditions for an inflation with the natural exit;
\item to find the initial values of a scalar field $\vphi$ and its rate of change $\dot{\vphi}$ which would be necessary and sufficient to begin an inflationary phase during the evolution of a scalar field;
\item to identify the effect that the ratio between the potential and kinetic terms of the energy density, as well as the initial values of $\vphi$, $\dot{\vphi}$, have on inflation phase, its time and on the number of e-folds;
\item to estimate the percentage of e-folds and the duration of the inflation over a period of a valid slow-rolling approximation.
\end{itemize}
%
%
%
%
\section{The relationship between the Friedman and Abel equations}
\label{Connection}
In this chapter we will follow the ideas of \cite{Yurov_Yurov} by introducing a transformation that would reduce the Friedman equations~(\ref{Friedman_H2}),~(\ref{Friedman_ddot_phi}) to the Abel equation of the 1st kind.\par
We will start by introducing a full energy density of scalar field $W$ (see \cite{Chervon_Zhuravlev_Shchigolev}):
\begin{equation}
W=\frac{1}{2}\dot{\vphi}^2+V(\vphi).
\label{W}
\end{equation}
Assuming that $k=0$ (i.e. concentrating on the case of a flat universe), we rewrite the system ~(\ref{Friedman_H2}),~(\ref{Friedman_ddot_phi}) as
\begin{equation}
\frac{dW}{d\vphi}=-3H\dot{\vphi},
\label{d_W_d_phi}
\end{equation}
\begin{equation}
H=\pm\frac{1}{\text{M}_{\text{P}}}\sqrt{\frac{8\pi}{3}W}.
\label{H_W}
\end{equation}
The knowledge of the function $W=W(\vphi)$ provides us with a way for explicit reconstruction of unknown quantities $\vphi=\vphi(t)$, $a=a(t)$, and $V=V(\vphi)$. If $W\neq0$, $\vphi(t)$ can be derived from the ordinary differential equation
\begin{equation}
\frac{d\vphi}{dt}=\mp\frac{\text{M}_{\text{P}}}{\sqrt{24\pi}}\frac{W'(\vphi)}{\sqrt{W(\vphi)}},
\label{d_phi_ot_t}
\end{equation}
while the exact form of potential $V$ will follow from:
\begin{equation}
V(\vphi)=W(\vphi)-\frac{1}{6}\frac{\text{M}_{\text{P}}^2}{8\pi}\frac{(W'(\vphi))^2}{W(\vphi)}\,.
\label{eq:V_ot_W}
\end{equation}
As for the scale factor $a(t)=\exp(\int\;H(t)\,dt)$, it can be derived by substitution of $\vphi(t)$ from (\ref{d_phi_ot_t}) into (\ref{H_W}) and the consequent integration.\par
An integration of (\ref{d_phi_ot_t}) gives rise to the constant $t_0$, which appears because of a translational invariance of Einstein--Friedman equations, and can be assimilated by a translation $t\rightarrow t+const$. In order for solution $\vphi=\vphi(t;t_0,C)$ to be general it should also contain a second independent constant $C$. This constant can be calculated if one considers (\ref{eq:V_ot_W}) as a differential equation w.r.t. variable $W(\vphi)$ with given $V(\vphi)$, for it is the general solution of this equation that would contain the thought after integration constant $C$. Thus, we can now formulate the following
\begin{proposition}
If for a given $V(\vphi)$ the general solution of equation (\ref{eq:V_ot_W}) is $W=W(\vphi,C)$, the general solution of the Friedman equations~(\ref{Friedman_H2}),~(\ref{Friedman_ddot_phi}) $\vphi(t;t_0,C)$ exactly corresponds to the general solution of (\ref{d_phi_ot_t}).
\end{proposition}
\begin{remark}
The case $W=0$ shall be considered separately. It is easy to verify that for any given $V\leq0$ one gets $H=0$ ($a(t)=a_0=const$ -- stationary universe), and that the general solution of ~(\ref{Friedman_H2}),~(\ref{Friedman_ddot_phi}) contains a single integration constant:
$$
\begin{cases}
\int\;\dfrac{d\vphi}{\sqrt{-2V(\vphi)}}=\pm(t_0). & V\neq0 \\
\vphi=\vphi_0                                       & V=0
\end{cases}
$$
This formula demands caution while choosing the sign of potential $V(\vphi)$. The obvious choice would be a non-positive function $V$ that would result in a real-valued integral (see \cite{Felder_Frolov_Kofman_Linde}). However, while it is true that the positive-valued potentials in general lead to the complex values of $\vphi$, it doesn't necessarily mean that only the non-positive potentials are physically significant. For example, despite the obvious positiveness of the potentials $V(\vphi) = \vphi^{2-1/(2n)}$, $\vphi>0$, $n \in \N=\{1,2,...\}$, all of them yield the real-valued $\vphi$.
\end{remark}
As we can see, the problem is now reduced to a task of finding the general solution of (\ref{eq:V_ot_W}) for a given $V(\vphi)$.\par
The following theorem provides a connection between the full energy density of the scalar field $W(\vphi,C)$, the potential $V(\vphi)$ and the solution of the Abel equation of 1st kind.
\begin{theorem}
Let $x=4\sqrt{3\pi}/\textrm{M}_{\textrm{P}}\,\vphi$, $\chi=\ln|V|$, $\kappa\pm1$. For a given $V(\vphi)$ the corresponding Hamiltonian $W=W(x,C)$ is defined as
\begin{equation}
W(x,C)=V(x) \left(1+\frac{1}{y^2-1}\right),
\label{W_V_y}
\end{equation}
where $y=y(x,C)\neq\pm1$ is a general solution of Abel equation of 1st kind:
\begin{equation}
y'=-\frac{1}{2}\left(y^2-1\right)\left(\kappa-\chi'y\right).
\label{eq:Abel_0}
\end{equation}
Moreover, the special case $V=0$ occurs if and only if $y=\pm1$ and hamiltonian $W$ has the form:
$$
W=Ce^{\kappa x}.
$$
\label{theor_1}
\end{theorem}
The proof of the theorem can be performed by a direct calculation.
\section{The $m^2\vphi^2/2$ model: inflation and slow-rolling condition}
The Abel representation (\ref{eq:Abel_0}) of the Friedman equations~(\ref{Friedman_H2}),~(\ref{Friedman_ddot_phi}) can be extremely useful even in those cases when one cannot find its exact solution. For example, let us consider the popular cosmological model with the quadratic potential
\begin{equation}
V(\vphi)=\frac{m^2\vphi^2}{2}.
\label{phi2}
\end{equation}
The corresponding Abel equation is non-integrable; however, it still allows for a rather effective way to study the possible occurrence of inflation and the natural exit from it.\par
\begin{figure}
\begin{center}
\includegraphics[width=.5\columnwidth]{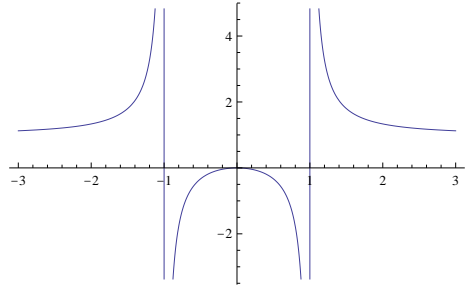}
\caption{\label{fig:theta} A graph of function $\theta(y)$. It can be seen that the range of $\theta$ is $(-\infty,0] \cup (1,+\infty)$.}
\end{center}
\end{figure}

Let us introduce the quantity: $\theta(y)$:
$$
\theta(y)=1+\frac{1}{y^2-1}.
$$
The graph of $\theta(y)$ is represented on (fig.~\ref{fig:theta}).

The definition (\ref{W}) together with (\ref{eq:Abel_0}) implies that
\begin{equation}
\frac{\dot\vphi^2}{2}=V(\theta-1).
\label{phi_theta}
\end{equation}
For the rest of this paper we will assume that $V \ge 0$ and $y \notin (-1,1)$, since this will guarantee that $\dot\vphi^2>0$. However, we would like to point out that the negative values of $\theta$ can not be regarded as entirely unphysical, for they correspond to the so called ``phantom fields'' -- a notion that has recently received a fair share of attention (see, for example, \cite{Dark-1,Dark-2,Dark-4,Dark-5,Dark-6,Yurov-3,ENO,Andrianov}).

The slow-rolling condition
$$
\frac{\dot{\vphi}^2}{2}\ll|V|
$$
can be rewritten as
\begin{equation}
\frac{V}{|V|} \left(\theta(y)-1\right) \ll 1.
\label{slow_rolling}
\end{equation}
Since (\ref{phi2}) requires $V \ge 0$, (\ref{slow_rolling}) is satisfied for any $|y| \gg \sqrt{2}$.\par
An inflation takes place whenever $\ddot{a}(t)/a(t)>0$. This condition is identical to inequality $\rho+3p<0$, which can in turn be reduced to $\theta(y)<3/2$ and, finally, to $|y| > \sqrt{3}$. The pressure $p$ will be negative if $\theta(y)<2$ or $|y|>\sqrt{2}$. Hence, we can determine the ranges for the values of $y$ that would correspond to the different stages of the universe's evolution (table~\ref{table_inf}).
\begin{table}[h]
\caption{}
\begin{tabular}{|c|c|c|c|}
\hline
 & Slow-rolling & Inflation & Negative pressure\\
\hline
I: $\ll y_*<y<\infty$ & yes & yes & yes \\
\hline
II: $\sqrt{3}<y<y_*$ & no & yes & yes \\
\hline
III: $\sqrt{2}<y<\sqrt{3}$ & no & no & yes \\
\hline
IV: $y<\sqrt{2}$ & no & no & no\\
\hline
\end{tabular}
\label{table_inf}
\end{table}
Thus, inflation might actually continue even after the slow-rolling condition has been compromised, because when $y$ crosses the range~II, the inflation would still be maintained up to the point when $y$ would finally reach $\sqrt{3}$.
\section{An application of the Abel equation of the 1st kind for analysis of the inflation in the model $m^2\vphi^2/2+\lambda\vphi^4/4$}
\subsection{The Abel equation of the 1st kind for the model $m^2\vphi^2/2+\lambda\vphi^4/4$}
Let us start by writing down the Abel equation of the 1st kind for a potential $m^2\vphi^2/2+\lambda\vphi^4/4$ in form (\ref{eq:Abel_0}). (As before, we restrict ourselves to the case $k=0$ for the flat universe.)\par
Following Theorem~\ref{theor_1}, after substitution
\begin{equation}
\vphi=\frac{x}{4\sqrt{3\pi}}\,\textrm{M}_{\textrm{P}}
\label{phi_x}
\end{equation}
we get a potential (\ref{V}) as a function of $x$:
\begin{equation}
V(x)=\frac{\lambda}{96\pi} x^2 \left(1+\frac{1}{96\pi} x^2\right)\,\textrm{M}_{\textrm{P}}^4,
\label{V(x)}
\end{equation}
assuming $m^2=\lambda\textrm{M}_{\textrm{P}}^2$, according to \cite{Linde}.
The Abel equation of the 1st kind, corresponding to the Friedman equations~(\ref{Friedman_H2}),~(\ref{Friedman_ddot_phi}), becomes
\begin{equation}
y'=-\frac{1}{2}\left(y^2-1\right)\left(1-\left(\frac{2}{x}+\frac{2x}
{96\pi+x^2}\right)y\right),
\label{eq:Abel_1}
\end{equation}
\par
The scalar field $\vphi$ is directly proportional to $x$. The connection between the rate of change of the scalar field $\dot{\vphi}$ and $x$, $\dot{x}$ and $y$ can be found from~(\ref{W}) after substitution of the expression (\ref{phi_x}).
\subsection{The initial conditions leading to inflation}
According to the table~\ref{table_inf} in the section~3, if we want the inflation to occur, the solution of the Abel equation~(\ref{eq:Abel_1}) has to be greater than $\sqrt{3}$ for a certain interval of values of $x$. Since the scalar field is proportional to $x$ and decreases during the inflation, the interval in question should belong to the interval $[0, x_0]$, where $x_0$ is the initial value of a scalar field.\par
\begin{figure}[h]
\begin{center}
\includegraphics[height=4.5cm,keepaspectratio,angle=-90]{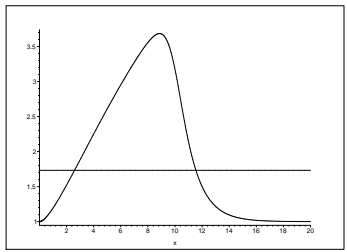}
\caption{The plot of the solution of the Abel equation~(\ref{eq:Abel_1}) for the initial values $y(12)=1{,}5$}
\label{Abel_resh}
\end{center}
\end{figure}
\begin{figure}[h]
\begin{center}
\includegraphics[height=4.5cm,keepaspectratio,angle=-90]{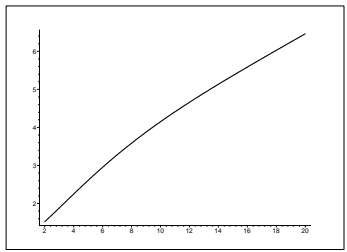}
\caption{The plot of maximal $y_0$, which provide finite solutions of the Abel equation~(\ref{eq:Abel_1}), for corresponding $x_0$}
\label{y0_max}
\end{center}
\end{figure}
\begin{figure}[h]
\begin{center}
\begin{minipage}[h]{0.49\linewidth}
\includegraphics[height=4.5cm,keepaspectratio,angle=-90]{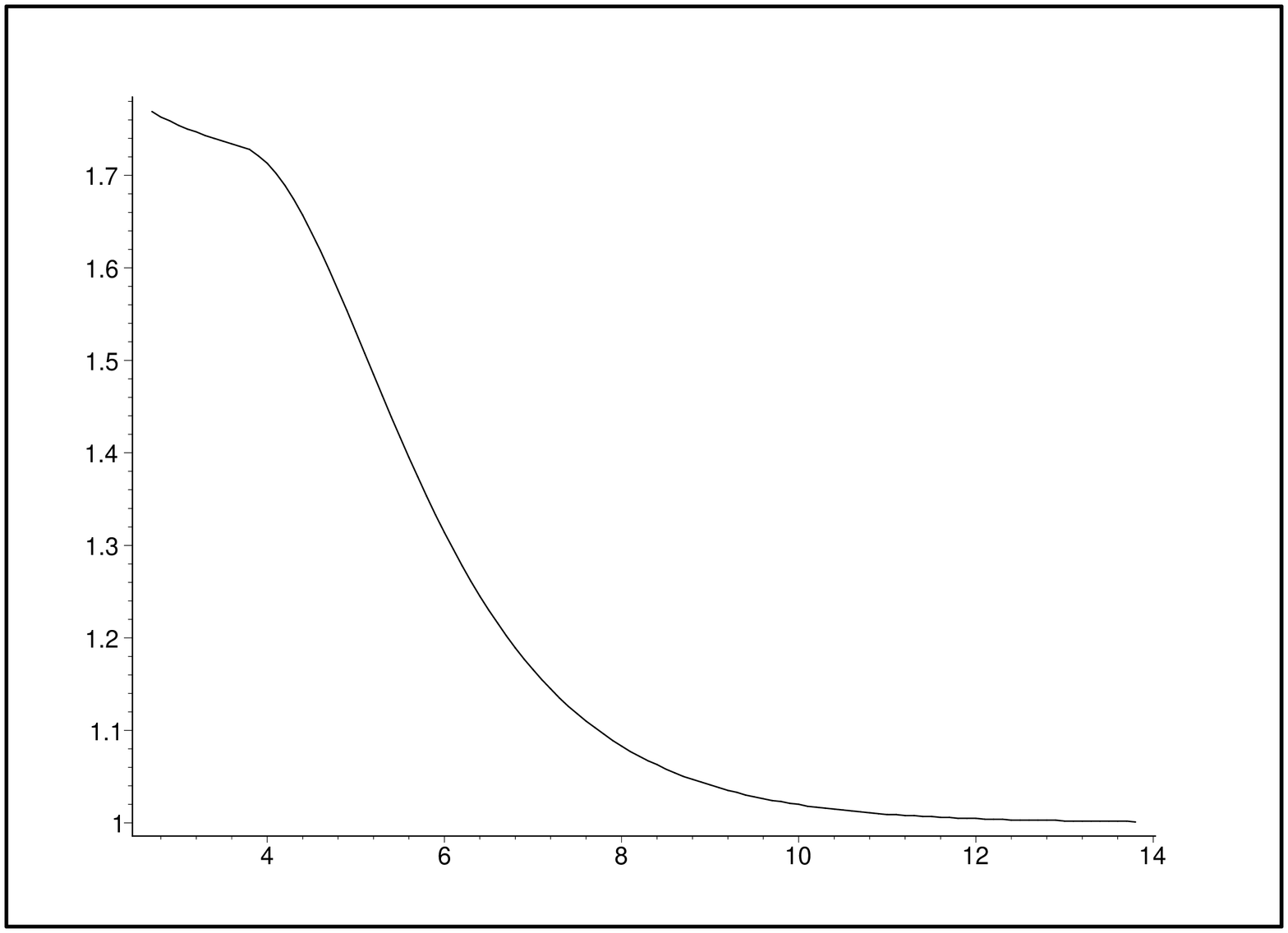}
\end{minipage}
\hfill
\begin{minipage}[h]{0.49\linewidth}
\includegraphics[height=4.5cm,keepaspectratio,angle=-90]{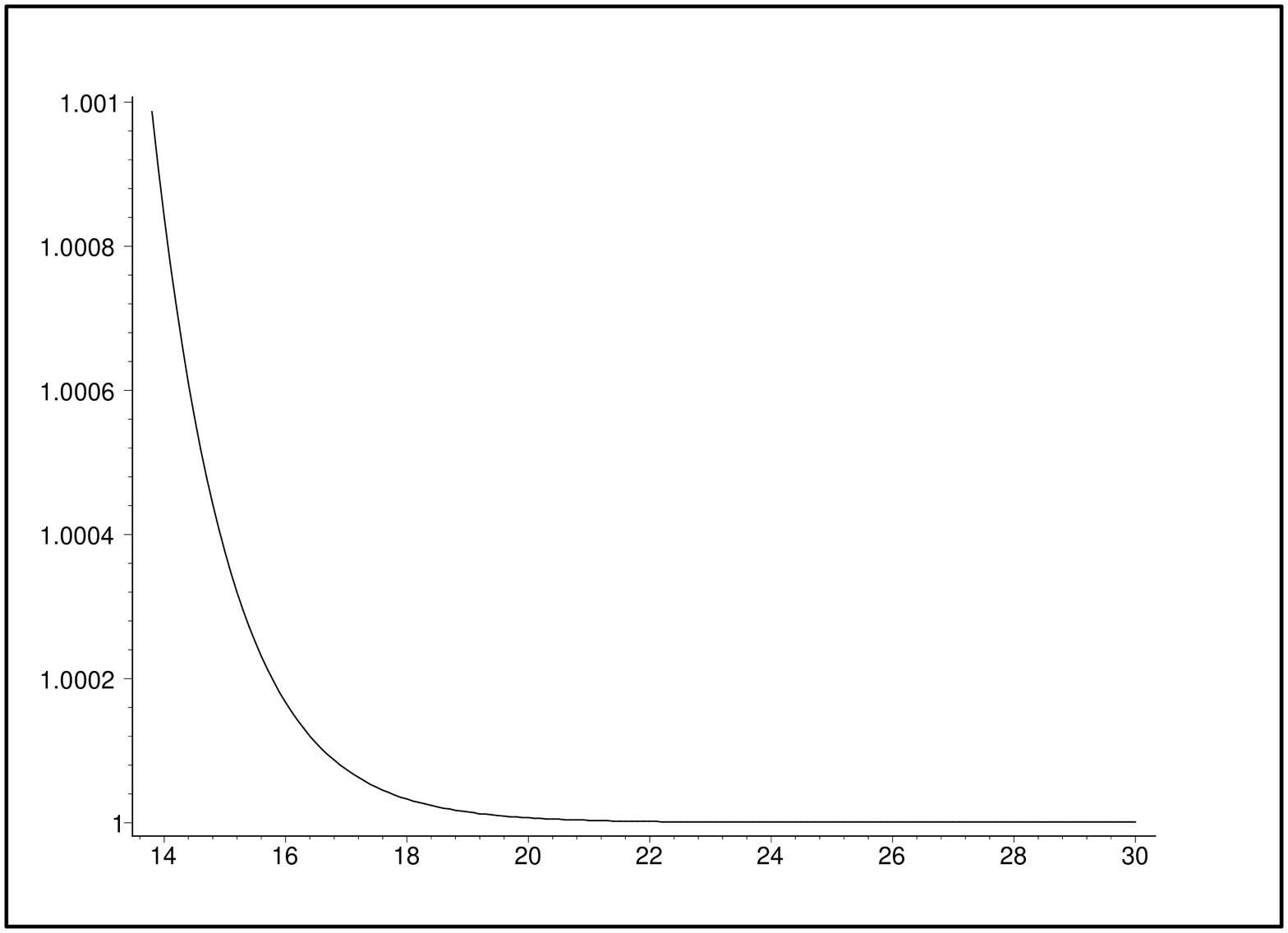}
\end{minipage}
\caption{The plot of minimal $y_0$, which provide finite solutions of the Abel equation~(\ref{eq:Abel_1}), and for which this solutions are more than $\sqrt{3}$ on some range of the $x$ axis, for corresponding $x_0$}
\label{y0_min}
\end{center}
\end{figure}

As a first step, the minimal initial values $y_0=y(x_0)$ were found, such that the corresponding solutions of the Abel equation~(\ref{eq:Abel_1}) (fig.~\ref{Abel_resh}) remained finite. \footnote{The singular solutions were not considered in this paper. The plot of maximal $y_0$ for corresponding $x_0$ still providing the finite solutions is on fig.~\ref{y0_max}. It is presumed that the singularity arises because of the violation of a Lipschitz condition \cite{Yurov_Yurov}}
The solutions of equation~(\ref{eq:Abel_1}) were derived by the computer algebra system Maple. The plots of $y_0$ for corresponding $x_0$ are displayed on the fig.~\ref{y0_min}. There one can see that an increase in $x_0$ leads to a decrease of a minimal $y_0$ necessary for inflation to occur. Up to $x_0=2{.}7$ there are no finite solutions of the equation~(\ref{eq:Abel_1}) which are more than $\sqrt{3}$ in the range of $x$ on the left of $x_0$, therefore, $x_0<2{.}7$ should be excluded from the further search for those initial conditions that would be sufficient for the inflation's occurrence.\par
\subsection{The sufficient initial conditions for inflation to occur}
Two parameters were chosen as the measures of the inflation: the logarithm of the ratio between the scale factor values at points $y(x_f)=\sqrt{3}$ ($x_f<x_i$) and $y(x_i)=y_i$ ($y_i\geq\sqrt{3}$, $x_i\leq x_0$), and the time interval separating these two moments. These parameters predicts an inflation provided the following estimates are true:
$$
\ln{\frac{a_f}{a_i}}\gtrsim100,
$$
$$
t_f-t_i\lesssim1{.}9\cdot10^8~\text{M}_{\text{P}}^{-1},
$$
where $a_i$, $t_i$ are the values of the scale factor and time when $y(x_i)=y_i$, and $a_f$, $t_f$ are, correspondingly, the value of the scale factor and the time when $y(x_f)=\sqrt{3}$.\par
To estimate $\ln\dfrac{a_f}{a_i}$ and ~$t_f-t_i$ ~we use the fact that $\dfrac{d(\ln a)}{dt}=H.$ Recalling~(\ref{W}), (\ref{H_W}), (\ref{W_V_y}), (\ref{V(x)}) yields
$$
d(\ln{a})=H(x)\,dt=H(x)\frac{dt}{dx}\,dx=\pm\frac{1}{6}\sqrt{\frac{W(x)}
{W(x)-V(x)}}
\,dx=-\frac{1}{6}y(x)\,dx.
$$
The choice of a minus sign is based on assumption that the universe is experiencing the inflation, thus making the Hubble parameter positive, and the rate of change of scalar field -- negative.

Next,
$$
\ln{\frac{a_f}{a_i}}=-\frac{1}{6}\int\limits_{x_i}^{x_f}y(x)\,
dx=\frac{1}{6}\int\limits_{x_f}^{x_i}y(x)\,dx,
$$
so, using ~(\ref{W}), (\ref{W_V_y}), (\ref{V(x)}), we have
$$
t_f-t_i=-\frac{\mathrm{M}_{\text{P}}}{4\sqrt{\pi}}\int\limits_{x_i}^{x_f}
\frac{dx}{\sqrt{W(x)-V(x)}}=
\frac{\mathrm{M}_{\text{P}}}{4\sqrt{\pi}}\int\limits_{x_f}^{x_i}
\sqrt{\frac{y^2(x)-1}{V(x)}}\,dx.
$$
The consequent numeric integration yields the required estimate.\par
\begin{figure}[h]
\begin{center}
\includegraphics[height=4.5cm,keepaspectratio,angle=-90]{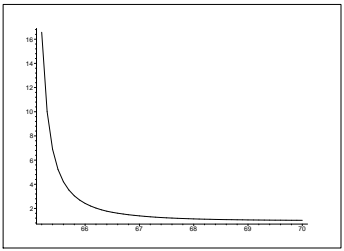}
\caption{The plot of a minimal $y_0(x_0)$, sufficient for an inflation to occur. ($\lambda=10^{-14}$)}
\label{inflation}
\end{center}
\end{figure}

To find the minimal values $(x_0,y_0)$, which are necessary for inflation to occur, we choose the following steps: $h_x=0{.}1$ for  $x_0$, $h_y=0{.}001$ for $y_0$. It is determined that inflation can occur if the initial values are more than $x_0=65{.}2$, $y_0=16{.}573$, which correspond to the scalar field value $\vphi_0=5{.}3\;\text{M}_{\text{P}}$ and its rate of change value $\dot{\vphi}_0=-1{.}2\cdot10^{-7}\;\text{M}_{\text{P}}^2$ for $\lambda=10^{14}$. In this case $\ln\dfrac{a_f}{a_i}$ and $t_f-t_i$ have the respective values $100{.}0$ and $1{.}1\cdot10^{8}\;\text{M}_{\text{P}}^{-1}$. During the inflationary phase $y_0$ decreases sharply along with the increase of $x_0$, just as plotted on fig.~\ref{inflation}.
\subsection{How the initial values $(x_0,y_0)$ affect the e-folds number and the time span of inflation}
\begin{figure}[t]
\begin{center}
\begin{minipage}[t]{0.49\linewidth}
\includegraphics[height=4.5cm,keepaspectratio,angle=-90]{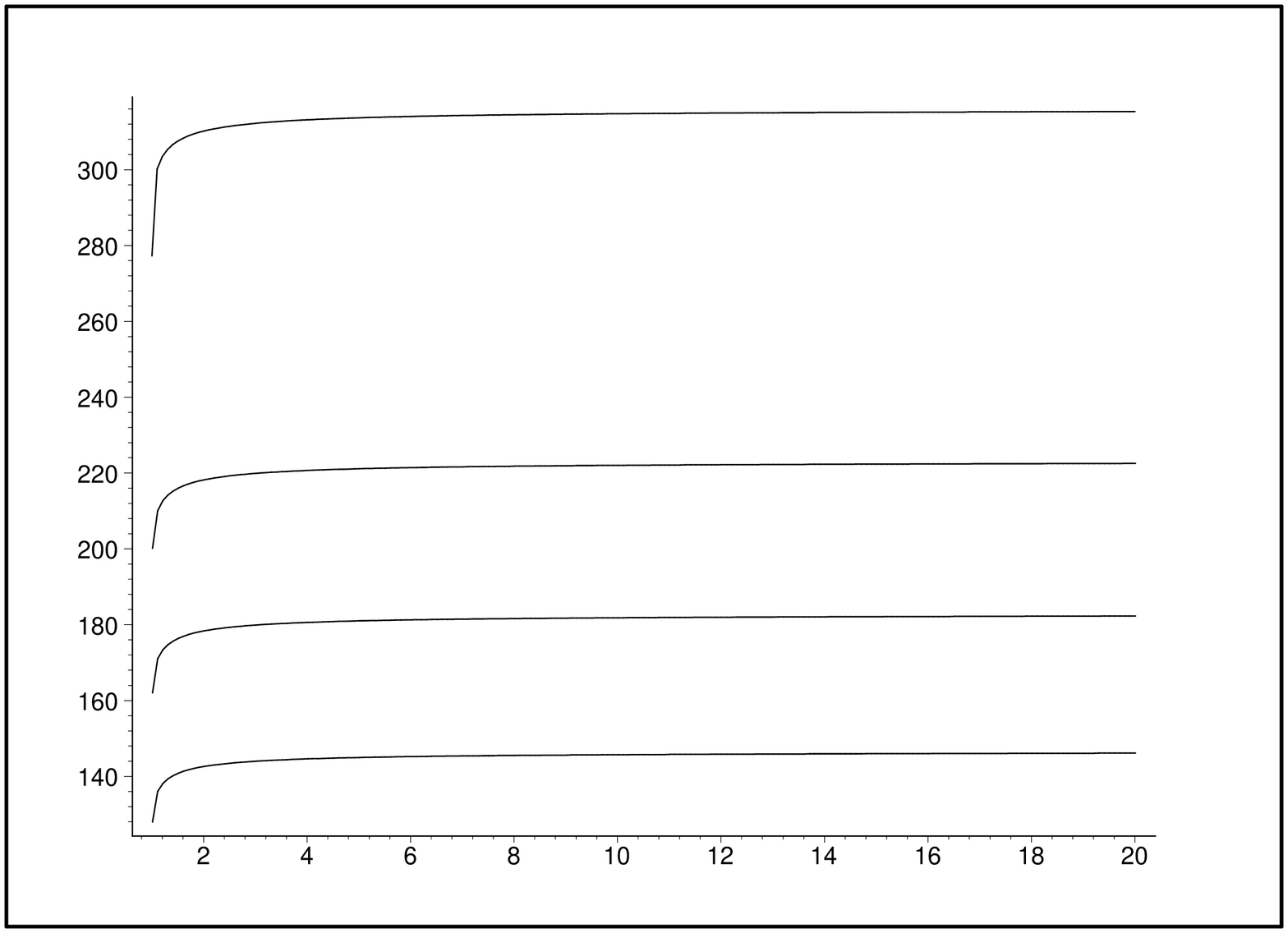}
\end{minipage}
\hfill
\begin{minipage}[t]{0.49\linewidth}
\includegraphics[height=4.5cm,keepaspectratio,angle=-90]{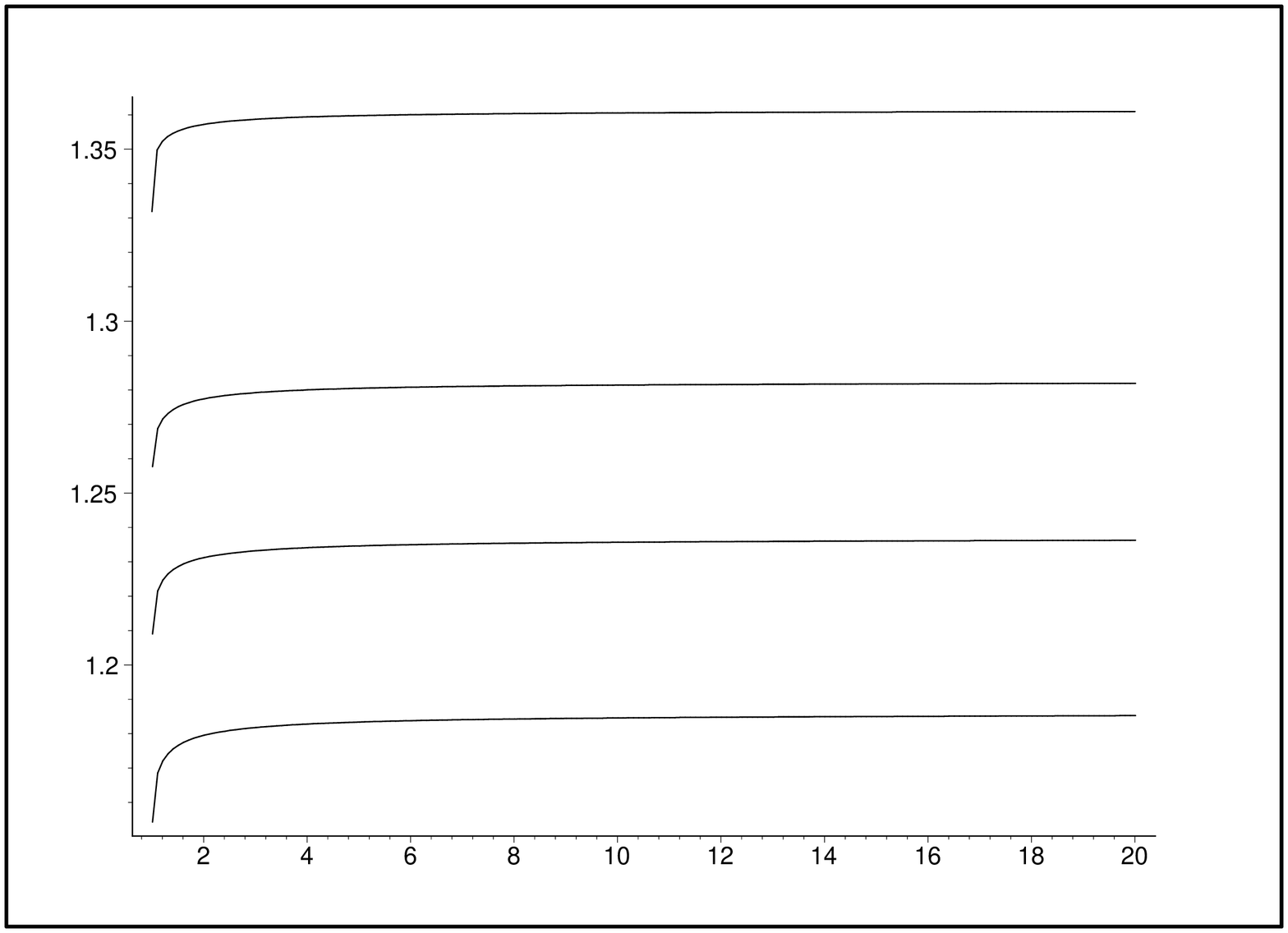}
\end{minipage}
\begin{minipage}[t]{1\linewidth}
\begin{tabular}{p{0.49\linewidth}p{0.49\linewidth}}
\centering\caption{The plot of the relation between the e-folds number and  $y_0$ for different $x_0$ (top-down $x_0=120$,\quad$x_0=100$,\quad$x_0=90$,\quad$x_0=80$)}
\label{ln(a)} & \centering\caption{The plot of the relation between the time of inflation  $t_{inf}\cdot10^{-8}\;\text{M}_{\text{P}}^{-1}$ and $y_0$ for different $x_0$ (top-down $x_0=120$,\quad$x_0=100$,\quad$x_0=90$,\quad$x_0=80$)}
\label{t_inf}
\end{tabular}
\end{minipage}
\end{center}
\end{figure}

The plots of the e-folds number and the inflationary time span are displayed on figs.~\ref{ln(a)}~and~\ref{t_inf} respectively.
One can notice that the e-folds number $P_{inf}={\ln\dfrac{a_f}{a_i}}$ and time period $t_{inf}=t_f-t_i$ grow fast when $y_0\gtrsim1$ and increases. However, the bigger $y_0$ gets the lesser its effect becomes. An increase of $x_0$ leads to a growth of both the e-folds number and inflationary time span. Also one can note that the dependence of the $P_{inf}$ and $t_{inf}$ with $y_0$ is smaller for greater $x_0$. The values of the e-folds number and the time span of inflation for different $x_0$, $y_0$ are shown in the tables~\ref{table1},~\ref{table2}.
\begin{table}[h]
\caption{}
\begin{tabular}{|c|c|c|c|c|}
\hline
$y_0$ & $\frac{P_{inf(x_0=400)}}{P_{inf(x_0=200)}}$ &
$\frac{t_{inf(x_0=400)}}{t_{inf(x_0=200)}}$ &
$\frac{P_{inf(x_0=600)}}{P_{inf(x_0=200)}}$ &
$\frac{t_{inf(x_0=600)}}{t_{inf(x_0=200)}}$\\
\hline
$10$ & $3{,}941$ & $1{,}19025$ & $8{,}840$ & $1{,}30152$\\
\hline
$20$ & $3{,}940$ & $1{,}19014$ & $8{,}834$ & $1{,}30138$\\
\hline
$30$ & $3{,}939$ & $1{,}19010$ & $8{,}831$ & $1{,}30132$\\
\hline
\end{tabular}
\label{table1}
\end{table}
\begin{table}[h]
\caption{}
\begin{tabular}{|c|c|c|c|c|}
\hline
$x_0$ & $\frac{P_{inf(y_0=20)}}{P_{inf(y_0=10)}}$ &
$\frac{t_{inf(y_0=20)}}{t_{inf(y_0=10)}}$ &
$\frac{P_{inf(y_0=30)}}{P_{inf(y_0=10)}}$ &
$\frac{t_{inf(y_0=30)}}{t_{inf(y_0=10)}}$\\
\hline
$200$ & $1{,}0011$ & $1{,}00015$ & $1{,}0015$ & $1{,}00021$\\
\hline
$400$ & $1{,}0005$ & $1{,}00006$ & $1{,}0007$ & $1{,}00008$\\
\hline
$600$ & $1{,}0004$ & $1{,}00004$ & $1{,}0005$ & $1{,}00005$\\
\hline
\end{tabular}
\label{table2}
\end{table}
They contain the ratio between the inflation time and e-folds number for $y_0=10$, $y_0=20$, $y_0=30$ and $x_0=200$,
$x_0=400$, $x_0=600$. From the table~\ref{table1} it appears that doubling and tripling of $x_0$ with $y_0$ fixed leads to the growth of the e-folds number which is more noticeable than that of an inflationary time span. The data in table~\ref{table2} shows that an increase in $y_0$ results in a slow growth of the e-folds number and a slower growth of the inflation's duration while $x_0$ is constant.
\subsection{How the scalar field and its rate of change depend on the $x$ and $y$ values}
\begin{figure}[h]
\begin{center}
\includegraphics[height=4.5cm,keepaspectratio,angle=-90]{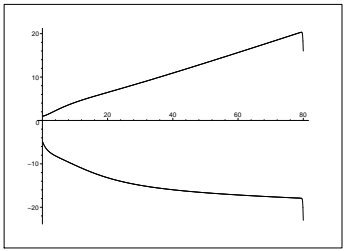}
\caption{The graphs of the solution of the Abel equation~(\ref{eq:Abel_1}) (upper curve) for the initial conditions
\text{$y(80)=16$} and the corresponding rate of change of the scalar field $\dot{\vphi}\cdot5\cdot10^{8}~\text{M}_{\text{P}}^2$ during its evolution (lower curve)}
\label{dot_phi}
\end{center}
\end{figure}

Expression (\ref{phi_x}) establishes a relationship between the scalar field and the $x$ value
$$
\vphi=\frac{x}{4\sqrt{3\pi}}\,\mathrm{M}_{\mathrm{P}}=8{,}1\cdot10^{-2}\cdot\,x~\text{M}_{\text{P}}.
\label{phi_ot_x}
$$
The corresponding relationship between the rate of change of a scalar field and the $x$ and $y$ variables can be obtained from the equations~(\ref{W}), (\ref{W_V_y}), (\ref{phi_x}), (\ref{V(x)})
\begin{equation}
\dot{\vphi}=-\frac{\sqrt{2\lambda}}{96\pi}\frac{x\sqrt{96\pi+x^2}}{\sqrt{y^2-1}}\,\mathrm{M}_{\mathrm{P}}^2=-4{,}7\cdot10^{-10}\cdot\frac{x\sqrt{96\pi+x^2}}{\sqrt{y^2-1}}\,\mathrm{M}_{\mathrm{P}}^2,
\label{dot_phi_ot_x_y}
\end{equation}
for $\lambda=10^{-14}$, $m^2=\lambda\textrm{M}_{\textrm{P}}^2$.
We use the values $y>1$ to resolve the equation~(\ref{eq:Abel_1}) in this case \cite{Yurov_Yurov}. We choose the minus sign, since during the evolution the rate of change of the scalar field has to be negative. The example of its evolution during the inflationary phase is displayed on the fig.~\ref{dot_phi}.\par
As a next step, let us turn out attention to the slow-rolling condition. More specifically, let us find out those $y$ for whom the slow-rolling condition breaks down:
$$
\frac{1}{2}\dot{\vphi}^2\ll|V(\vphi)|
$$
or
\begin{equation}
\frac{\dot{\vphi}^2}{2|V(\vphi)|}\ll1.
\label{slow_rolling_1}
\end{equation}
Under the assumption that $\dfrac{1}{2}\dot{\vphi}^2$ has to be at least ten times smaller than $|V(\vphi)|$, we determine that the slow-rolling condition is violated when $y\lesssim\sqrt{11}$.\par
Now, what can be said about an influence that the initial value of the scalar field $\vphi_0$ and initial ratio $2\dfrac{|V(\vphi_0)|}{\dot{\vphi}_0^2}$ have on the e-folds number and the inflation time? From the expression~(\ref{phi_ot_x}) we get
\begin{equation}
x_0=\frac{4\sqrt{3\pi}}{\mathrm{M}_{\mathrm{P}}}\vphi_0,
\label{x0_ot_phi_0}
\end{equation}
and from (\ref{slow_rolling_1})
\begin{equation}
y_0=\sqrt{2\frac{V(\vphi_0)}{\dot{\vphi}_0^2}+1}.
\label{y0}
\end{equation}
In the subsection~4.4 we have shown that the increase in $x_0$ and $y_0$ leads to a growth of the e-folds number and inflation time, and  the $y_0$ effect is insignificant when $y_0\gtrsim2$. Moreover, $y_0$, which is necessary for inflation to occur, decreases when $x_0$ grows. Thus, from the expressions~(\ref{x0_ot_phi_0}),
~(\ref{y0}) one can draw the conclusion that the inflation occurrence, the e-folds number and the time of inflation all have a strong dependence on the initial value of the scaler field $\vphi_0$ and a weak dependence on the initial ratio between the potential and kinetic terms of the energy density of the scalar field. The e-folds number and inflation time increase with the initial value of the scalar field rising, according to the expression~(\ref{x0_ot_phi_0}). An increase in the initial rate of change results in a decrease in $y_0$, and the e-folds number and inflation time diminish slowly. If $|\dot{\vphi_0}|$ is big enough, $y_0$ becomes of the order of $1$ and $P_{inf}$ and $t_{inf}$ decrease considerably. Thus, the choice of the initial value of a scalar field has a strong impact on both the beginning and intensity of the inflationary phase, while the impact of the initial {\em rate of change} of $\varphi$ remains comparatively little until $2|V(\vphi_0)|/\vphi_0^2\gtrsim3$. So, $V(\vphi_0)$ grows while $\vphi_0$ increases, and after that bigger and bigger values of $|\dot{\vphi_0}|$ becomes necessary for a violation of this condition and an abrupt decrease of $P_{inf}$ and $t_{inf}$.\par
The natural exit from inflation occurs when $y<\sqrt{3}$, as shown in \cite{Yurov_Yurov}.
\subsection{How the initial values of a scalar field and the energy ratio affect the slow-rolling condition and the process of inflation.}
\begin{figure}[h]
\begin{center}
\begin{minipage}[h]{0.49\linewidth}
\includegraphics[height=4.5cm,keepaspectratio,angle=-90]{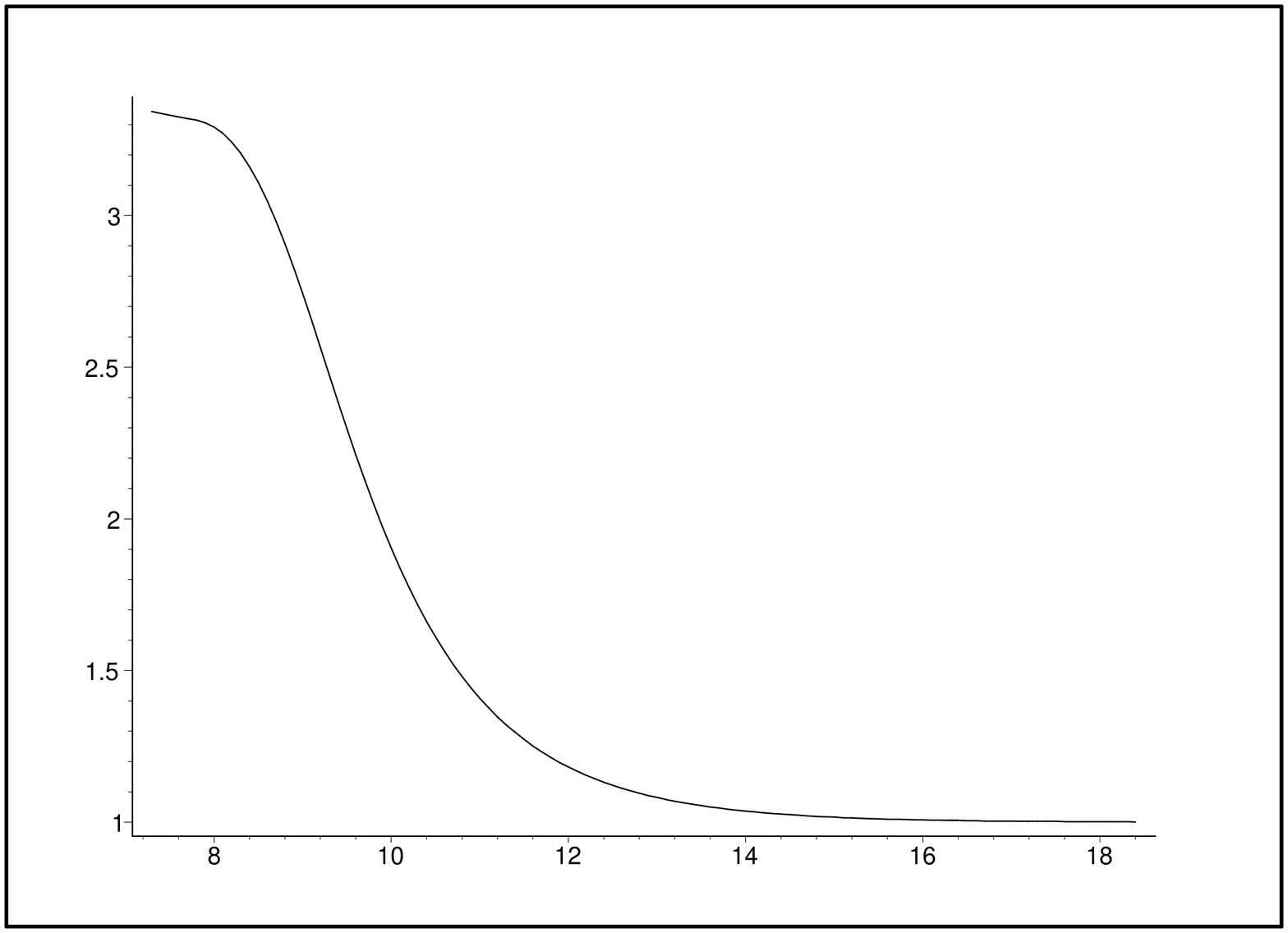}
\end{minipage}
\hfill
\begin{minipage}[h]{0.49\linewidth}
\includegraphics[height=4.5cm,keepaspectratio,angle=-90]{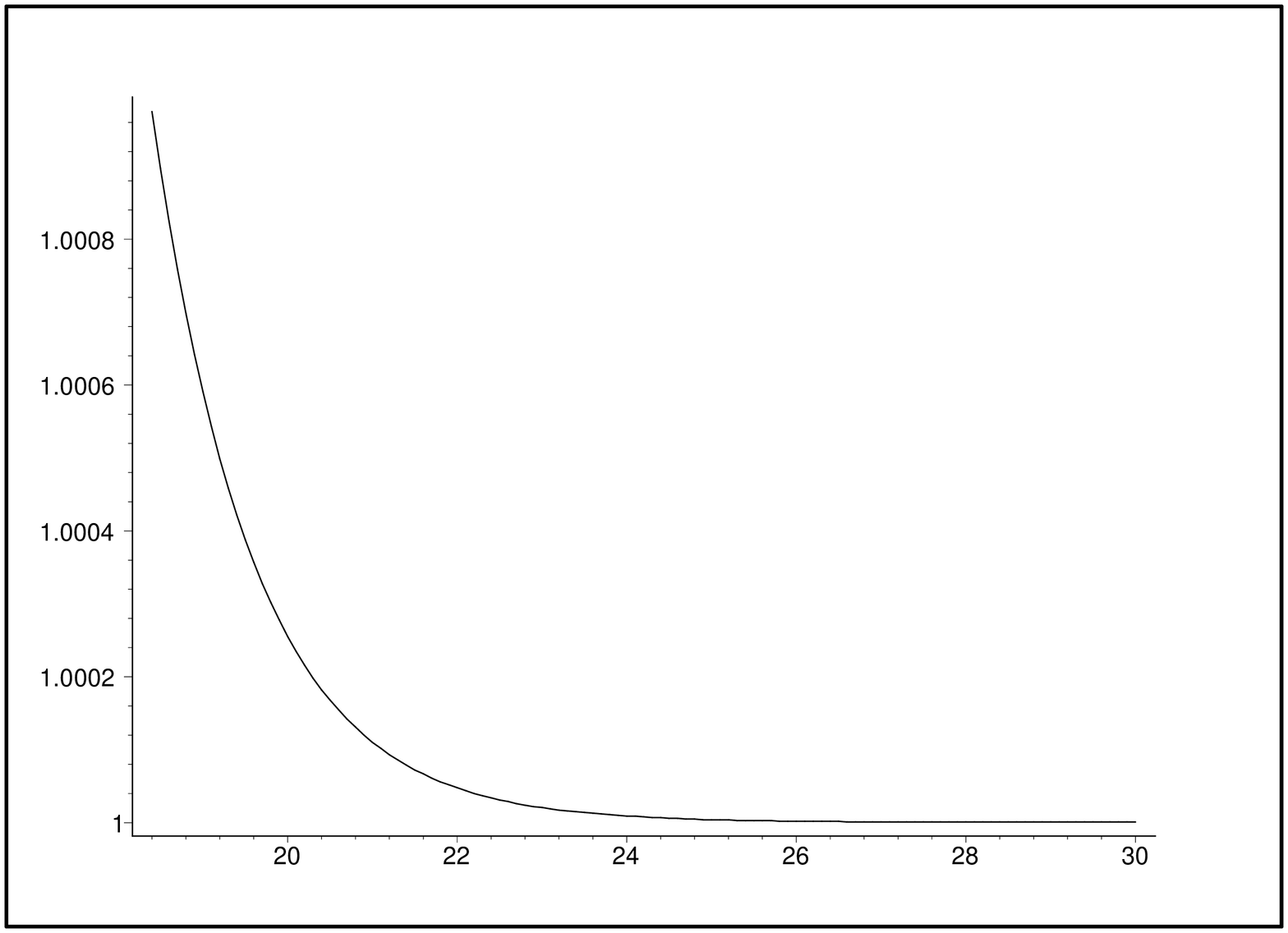}
\end{minipage}
\caption{The plot of minimal $y_0$, which provide finite solutions of the Abel equation~(\ref{eq:Abel_1}), and for which this solutions are more than $\sqrt{11}$ on some range of the $x$ axis, for corresponding $x_0$}
\label{sqrt(11)}
\end{center}
\end{figure}

\begin{figure}[h]
\begin{center}
\includegraphics[height=4.5cm,keepaspectratio,angle=-90]{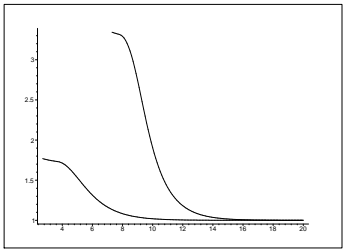}
\caption{The plots of the minimal $y_0(x_0)$ that (on some interval of $x$ axis) yields the finite solutions of Abel equation~(\ref{eq:Abel_1}) bigger than $\sqrt{3}$  (lower curve) and $\sqrt{11}$ (upper curve), correspondingly.}
\label{3_11}
\end{center}
\end{figure}

It is a safe claim that, should the scalar field get big enough, the slow-rolling condition would follow regardless of an initial ratio between the potential and kinetic terms of its energy density. In other words, the bigger the scalar field gets, the smaller an influence of the ratio ends up being. For example, for $x_0=100$ (which corresponds to $\vphi_0=8{.}05\;\text{M}_{\text{P}}$) the slow-rolling condition fails when $2|V(\vphi_0)|/\dot{\vphi}_0^2\lesssim10^{-18}$, and for $x_0=200$ ($\vphi_0=16{.}1\;\text{M}_{\text{P}}$) the failure occurs only when $2|V(\vphi_0)|/\dot{\vphi}_0^2\lesssim10^{-21}$ for $\lambda=10^{-14}$. The plot on the fig.~\ref{sqrt(11)} shows minimal $y_0$ for corresponding $x_0$, which is necessary for solution of Abel equation~(\ref{eq:Abel_1}) to be more than $\sqrt{11}$ on some range of values of $x$ located on the left of $x_0$.
These values satisfy the slow-rolling condition. The fig.~\ref{3_11} combines two plots of $y_0$ for corresponding $x_0$, for which solutions of equation~(\ref{eq:Abel_1}) exceed $\sqrt{3}$ and $\sqrt{11}$.
From the figures one can notice that $y_0$ required for the slow-rolling condition to occur diminishes as $x_0$ grows, i.e. the bigger initial value of the scalar field imposes less restrictions on its initial rate of change.\par
However, although for $\lambda=10^{-14}$ the slow rolling condition occurs when $\vphi_0=0{.}6\;\text{M}_{\text{P}}$ and $\dot{\vphi}_0=-2{.}0\cdot 10^{-8}\;\text{M}_{\text{P}}^2$, the condition $\ln\dfrac{a_f}{a_i}\gtrsim100$ requires $\vphi_0=5{.}3\;\text{M}_{\text{P}}$ and $\dot{\vphi}_0=-1{.}2\cdot 10^{-7}\;\text{M}_{\text{P}}^2$, so even if the slow-rolling condition rises but $\vphi_0<5{.}3\;\text{M}_{\text{P}}$, we will only have a ``mild'' version of inflation with $\ln\dfrac{a_f}{a_i}\lesssim100$.\par

\begin{figure}[h]
\begin{center}
\includegraphics[height=4.5cm,keepaspectratio,angle=-90]{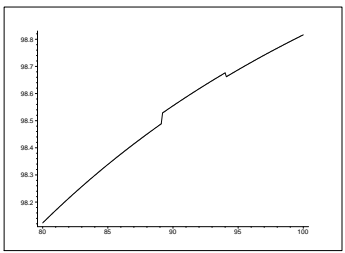}
\caption{The percentage of a number of e-folds arising during the slow-rolling phase as a function of $x_0$ for $y_0=1{,}001$.}
\label{a1}
\end{center}
\end{figure}

\begin{figure}
\begin{center}
\includegraphics[height=4.5cm,keepaspectratio,angle=-90]{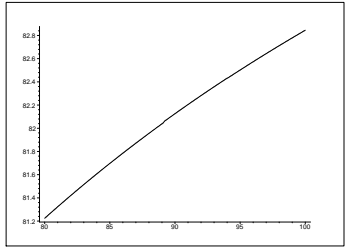}
\caption{The percentage of inflation's time spent during the slow-rolling phase as a function of $x_0$ for $y_0=1{,}001$.}
\label{t1}
\end{center}
\end{figure}

\begin{figure}[h]
\begin{center}
\begin{minipage}[h]{0.49\linewidth}
\includegraphics[height=4.5cm,keepaspectratio,angle=-90]{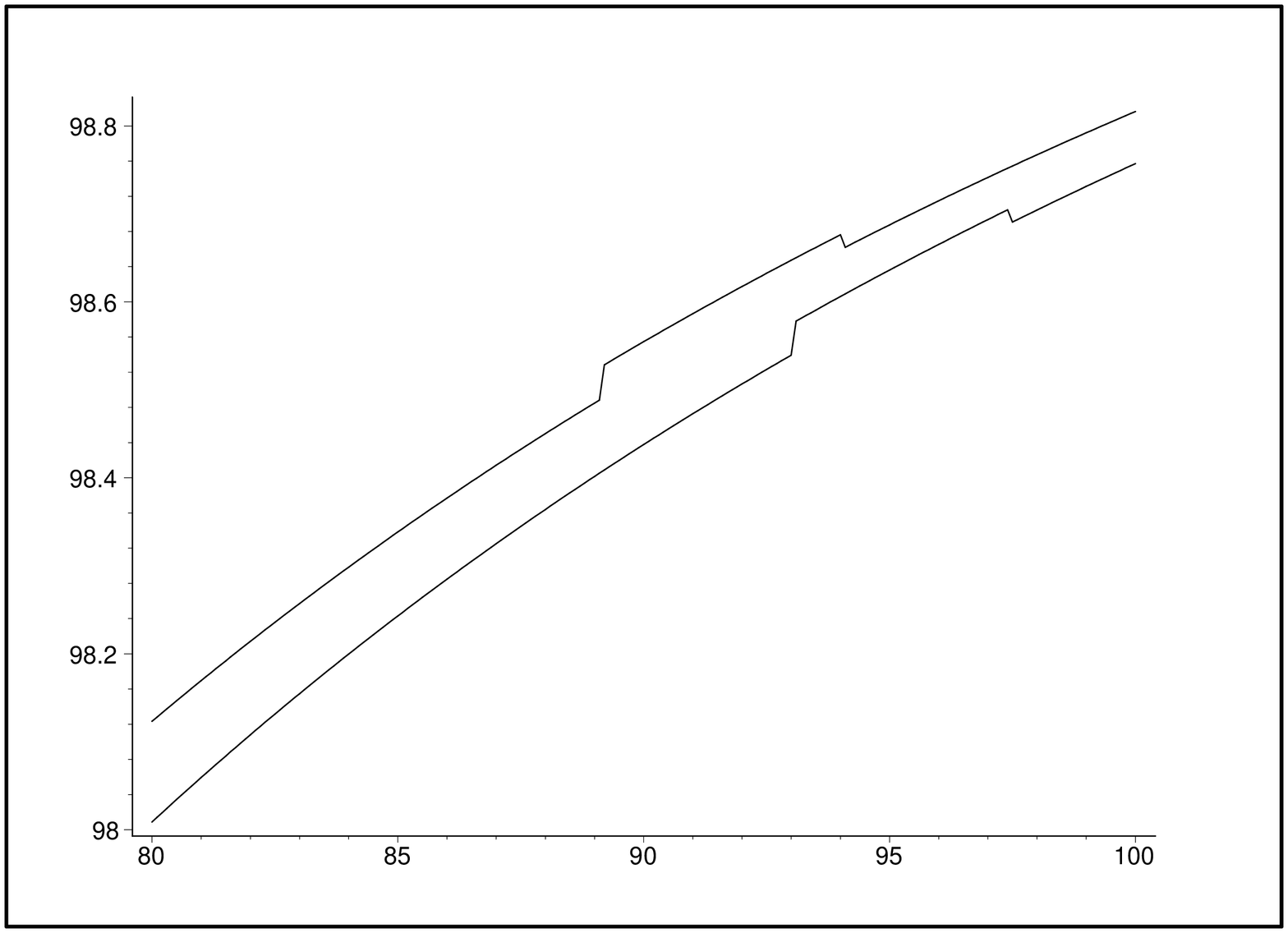}
\end{minipage}
\hfill
\begin{minipage}[h]{0.49\linewidth}
\includegraphics[height=4.5cm,keepaspectratio,angle=-90]{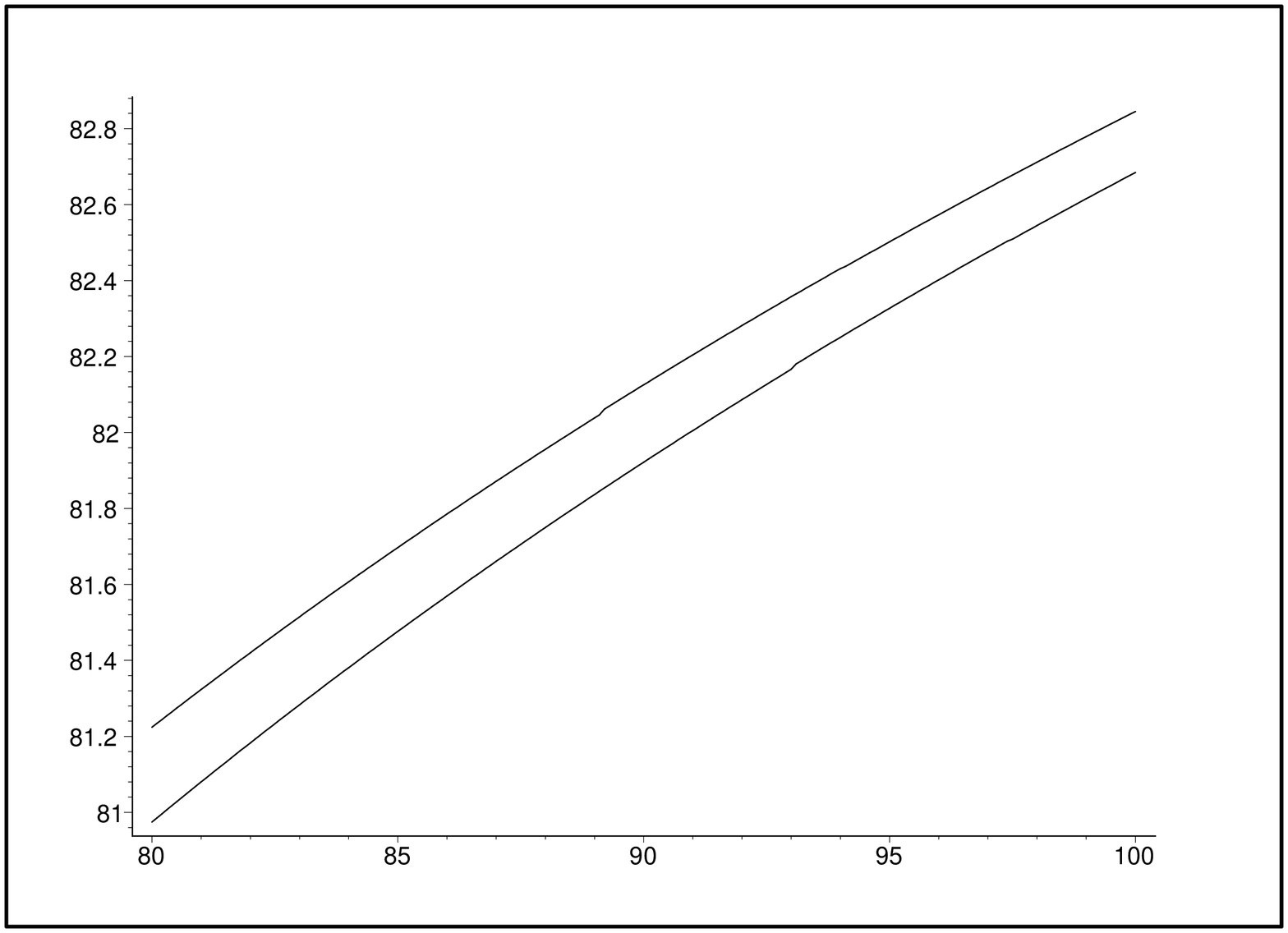}
\end{minipage}
\caption{The percentages of the e-folds number (left) and duration (right) of the slow-rolling phase as functions of $x_0$ for $y_0=1{,}25125$ (upper curve) and $y_0=1{,}001$ (lower curve)}
\label{a_sr_t_sr}
\end{center}
\end{figure}

In order to estimate the influence of the slow-rolling condition on inflation for different $x_0$, we have plotted the graphs of two relations: first, the one involving the percentage of all e-folds happening during the slow-rolling phase (as opposed to the total number of e-folds that is due to inflation in general), and the other one of the percentage of inflationary time spent during the slow-rolling phase. Both graphs are plotted as functions of $x_0$ for the value $y_0=0{,}001$ (fig.~\ref{a1}~and~fig.~\ref{t1} respectively).
On the figures one can see that the slow-rolling condition is indeed satisfied during the bigger part of an inflationary phase. As a matter of fact, it lasts for approximately $82\%$ of the inflationary phase and accounts for as much as $98\%$ of the total e-folds number. If the initial value of the scalar field increases while the initial ratio between the potential and kinetic terms remains constant, those numbers gets even bigger, although the actual growth during the inflation slows down. On the other hand, a comparable increase of the initial ratio between the potential and kinetic terms (with the constant initial value of the scalar field), would leave the percentages virtually unscathed, yielding only a minor increase in numbers (fig.~\ref{a_sr_t_sr}).
\subsection{The comparative study of $m^2\vphi^2/2+\lambda\vphi^4/4$, $m^2\vphi^2/2$ and $\lambda\vphi^4/4$ models}
\begin{figure}[h]
\begin{center}
\includegraphics[height=4.5cm,keepaspectratio,angle=-90]{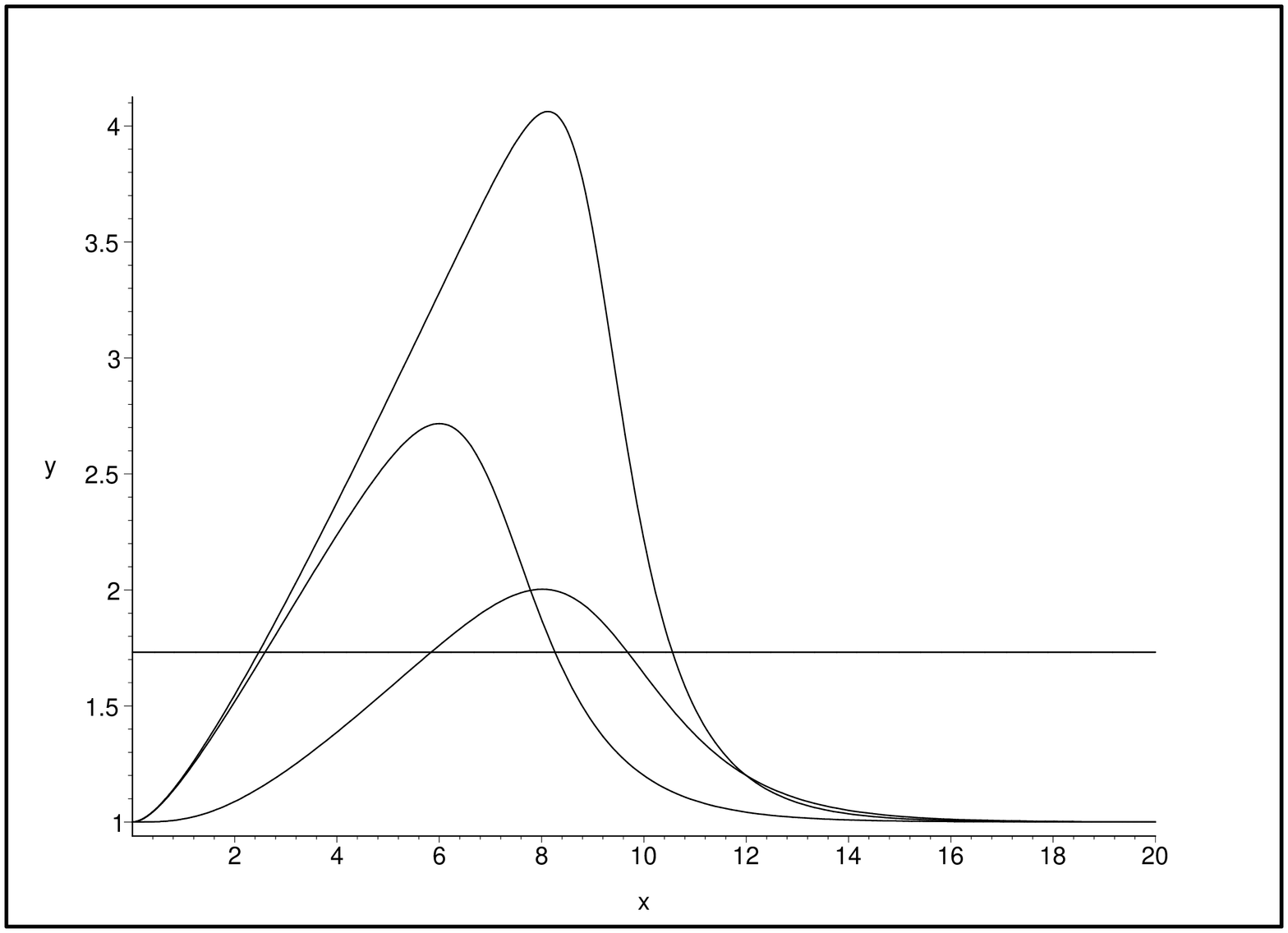}
\caption{The plots of the solutions of the Abel equations, corresponding to models
$m^2\vphi^2/2+\lambda\vphi^4/4$ (center curve),
$m^2\vphi^2/2$ (upper curve), $\lambda\vphi^4/4$
(lower curve), for the initial value $y(12)=1{,}5$}
\label{y(x)_multi}
\end{center}
\end{figure}

For a finishing touch, let us compare the inflationary dynamics of three models: $m^2\vphi^2/2+\lambda\vphi^4/4$, $m^2\vphi^2/2$ and $\lambda\vphi^4/4$ (table~\ref{comparison}).
\begin{table}[h]
\caption{The comparison of the results for $m^2\vphi^2/2+\lambda\vphi^4/4$, $m^2\vphi^2/2$ and $\lambda\vphi^4/4$ models with $\lambda=10^{-14}$, $m^2=\lambda\mathrm{M}_{\mathrm{P}}^2$}
\begin{tabular}{|c|c|c|c|}
\hline
$V(\vphi)$ & $\dfrac{m^2\vphi^2}{2}+\dfrac{\lambda\vphi^4}{4}$ &
$\dfrac{m^2\vphi^2}{2}$ &
$\dfrac{\lambda\vphi^4}{4}$\\[5pt]
\hline
$V(x)$ &
$\dfrac{\lambda}{96\pi}x^2(1+\dfrac{1}{96\pi}x^2)\,\mathrm{M}_{\mathrm{P}}^4$ &
$\dfrac{\lambda}{96\pi}x^2\,\mathrm{M}_{\mathrm{P}}^4$ &
$\dfrac{\lambda}{9216\pi^2}x^4\,\mathrm{M}_{\mathrm{P}}^4$\\[5pt]
\hline
\multicolumn{4}{|c|}{Minimal $\vphi_0$ and $\dot{\vphi}_0$, sufficient for inflation to occur}\\
\hline
$x_0$ &
$62{.}5$ &
$48{.}8$ &
$69{.}2$\\
\hline
$y_0$ &
$16{.}573$ &
$21{.}615$ &
$16{.}694$\\
\hline
$\vphi_0,~\text{M}_{\text{P}}$ &
$5{.}3$ &
$4{.}0$ &
$5{.}6$\\
\hline
$\dot{\vphi}_0,~\text{M}_{\text{P}}^2$ &
$-1{.}2\cdot10^{-7}$ &
$-1{.}8\cdot10^{-8}$ &
$-1{.}3\cdot10^{-7}$\\
\hline
E-folds number& $100{.}0$ & $100{.}0$ & $100{.}0$\\
\hline
Inflation time& $1{.}1\cdot10^8$ & $2{.}2\cdot10^8$ &
$1{.}1\cdot10^8$\\
$t_{inf},~\text{M}_{\text{P}}^{-1}$ & & &\\
\hline
\multicolumn{4}{|c|}{Minimal requirements on $\vphi_0$ and $\dot{\vphi}_0$ for the slow-rolling phase}\\
\hline
$x_0$ &
$7{.}3$ &
$6{.}2$ &
$12{.}8$\\
\hline
$y_0$ &
$3{.}343$ &
$3{.}361$ &
$3{.}339$\\
\hline
$\vphi_0,~\text{M}_{\text{P}}$ &
$5{.}3$ &
$4{.}0$ &
$5{.}6$\\
\hline
$\vphi_0,~\text{M}_{\text{P}}$ &
$0{.}6$ &
$0{.}5$ &
$1{.}0$\\
\hline
$\dot{\vphi}_0~\text{M}_{\text{P}}^2$ &
$-2{.}0\cdot10^{-8}$ &
$-1{.}6\cdot10^{-8}$ &
$-2{.}4\cdot10^{-8}$\\
\hline
\end{tabular}
\label{comparison}
\end{table}
The Abel equations describing the two latter models are:
\begin{equation}
y'=-\frac{1}{2}(y^2-1)\left(1-\frac{2y}{x}\right),
\label{eq:Abel_2}
\end{equation}
\begin{equation}
y'=-\frac{1}{2}(y^2-1)\left(1-\frac{4y}{x}\right).
\label{eq:Abel_4}
\end{equation}
The appropriate graphs of solutions of equations~(\ref{eq:Abel_1}),~(\ref{eq:Abel_2})~and~(\ref{eq:Abel_4}) can be seen on fig.~\ref{y(x)_multi}.\par
Table~\ref{comparison} shows that the inflationary dynamics in $m^2\vphi^2/2+\lambda\vphi^4/4$ and $\lambda\vphi^4/4$ impose the lower bounds on initial values that are actually much higher than in $m^2\vphi^2/2$ case, thanks to the fact that for $\vphi \gg 1$ the scalar field of the latter changes much slower than it does for the first two. Moreover, it appears that the slow-rolling stage in $m^2\vphi^2/2$ accounts for considerably bigger chunk of the inflationary phase than it does in the other two cases. The same is true for the e-folds number. Hence, although all three models yield the same e-folds number, the initial value of the scalar field for the $m^2\vphi^2/2$ potential ends up being the smallest of the three, while the duration of inflation happens to be the longest. The models $m^2\vphi^2/2+\lambda\vphi^4/4$ and $\lambda\vphi^4/4$ are quite similar to each other, but in the first case the e-folds number associated with the slow-rolling condition is bigger, and the minimal $\vphi_0$ necessary for inflation to occur is smaller than in the second case.\par
As for the minimal initial rate of change of the scalar field, it is only natural for its absolute value to decrease along with the initial value of the scalar field, since the small value of $2|V(\vphi_0)|/\dot{\vphi_0^2}$ is more favorable for the inflation per se.
\section{Conclusion}
In this paper we have shown that the existing connection between the Einstein--Friedman equations and the Abel equation can be successfully used as a primary means of analysis of many important cosmological models. In particular, we have applied it to the study of the inflationary dynamics in the models describing the flat homogenous isotropic universe filled with scalar field $\vphi$ with the potential $m^2\vphi^2/2+\lambda\vphi^4/4$. It has been found that:
\begin{itemize}
\item in most models the slow-rolling condition arises naturally during the dynamics of the scalar field; the necessary criterion for its absence is the small initial ratio between the potential and kinetic terms of the scalar field energy density; the growth of the initial value of the scalar field leads to a decrease of a necessary ratio; for example, if $\vphi_0=8{.}5\;\text{M}_{\text{P}}$ then the slow-rolling condition cannot occur when $2|V(\vphi_0)|/\dot{\vphi}_0^2\lesssim10^{-18}$;
\item the e-folds number and the time span of inflation both grow with the increase in the initial value of the scalar field and/or initial ratio between the potential and kinetic terms, although the time span grows noticeably slower; however, the main influence on the process of the inflation has to be attributed to the initial value of the scalar field;
\item if $|\dot{\vphi}_0|$ increases while $\vphi_0$ is constant then the e-folds number and the time of inflation decrease, however, if $\vphi_0$ is big enough, the change would be negligible; the effect becomes more noticeable if $|\dot{\vphi}_0|$ is sufficiently large and $2|V(\vphi_0)|/\dot{\vphi}_0^2\lesssim3$;
\item the ratio $2|V(\vphi_0)|/\dot{\vphi}_0^2$ has little influence on the process of inflation while $2|V(\vphi_0)|/\dot{\vphi}_0^2\gtrsim3$; thus, if $\vphi_0$ is big enough, the restriction on $\vphi_0$ can be relaxed, and as $\vphi_0$ grows, so does the highest possible value of $|\dot{\vphi}_0|$;
\item the condition $2|V(\vphi)|/\dot{\vphi}^2>2$ is necessary to initiate the inflation, while the natural exit from inflation requires $2|V(\vphi)|/\dot{\vphi}^2<2$, i.e. the inflation begins well before the slow-rolling phase and ends some time after;
\item the bigger part of the e-folds number and the time span of the inflation falls on the period of the slow-rolling condition (about $98\%$ for the e-folds and $82\%$ for the time); these percentages grow with the initial value of the scalar field and the initial ratio between the potential and kinetic terms of its energy density, although the former imposes bigger influence than the latter.
\end{itemize}
The results are on accord with the earlier estimates \cite{Linde}. This confirms the reliability of the analysis method suggested in \cite{Yurov_Yurov} and shows its potential for usage in other, more complicated models.
\newpage

\end{document}